# High Sensitivity Hybrid PbS CQD - TMDC Photodetectors up to 2 μm

Onur Özdemir,† Iñigo Ramiro,† Shuchi Gupta† and Gerasimos Konstantatos*,†,‡

†ICFO – Institut de Ciències Fotòniques, Av. Carl Friedrich Gauss 3, 08860 Castelldefels (Barcelona), Spain
‡ICREA – Institucio Catalana de Recerca i Estudis Avançats, Passeig Lluís Companys 23, 08010 Barcelona, Spain

## ABSTRACT

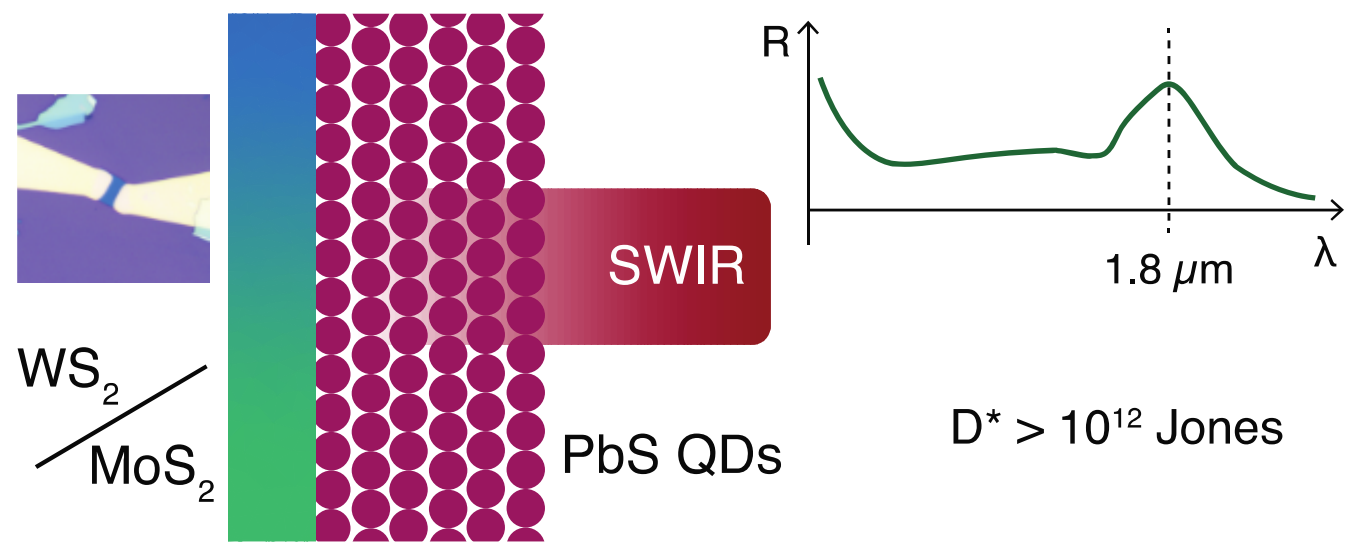

Recent approaches to develop infrared photodetectors characterized by high sensitivities, broadband spectral coverage, easy integration with silicon electronics and low cost have been based on hybrid structures of transition metal dichalcogenides (TMDCs) and PbS colloidal quantum dots (CQDs). However, to date, such photodetectors have been reported with high sensitivity up to 1.5 μm. Here we extend the spectral coverage of this technology towards 2 μm demonstrating for the first time compelling performance with responsivities 1400 A/W at 1.8 μm with 1V bias and detectivities as high as $10^{12}$ Jones at room temperature. To do this we studied two TMDC materials as a carrier transport layer, tungsten disulfide ($WS_2$) and molybdenum disulfide ($MoS_2$) and demonstrate that $WS_2$ based hybrid photodetectors outperform those of $MoS_2$ due to a more adequate band alignment that favors carrier transfer from the CQDs.

Photodetection in the infrared is of paramount importance for applications in night-vision, medical diagnosis, astronomy, environmental pollution monitoring, spectroscopy and telecommunications.[1–3] However, commercial infrared detectors are costly due to their epitaxial growth methods and complex non-monolithic integration with CMOS technology. Moreover, their performance is optimized upon cooling adding further complexity on their integration, power consumption and miniaturization.[4]

Colloidal quantum dots (CQDs) provide an attractive low-cost approach to traditional bulk, epitaxially grown materials generally used in photodetection in the infrared, such as InSb, InGaAs, HgCdTe.[5] They can be synthesized and processed in large quantities by solution processed techniques, with controllable size distributions. One of the most extensively studied materials employed in photodetectors has been CQDs of lead sulfide (PbS), that offers bandgap tunability from 2300 nm down to 700 nm.[6]

In CQD based photodetectors, a major obstacle against high performance is the low mobility of the quantum dot (QD) films, mainly due to inefficient charge hopping transport.[7] Surface engineering and ligand exchange strategies have been able to push mobilities up to 0.1 $cm^2/Vs$ in CQD films.[8] In some QD systems, developing composition-matched contacts for the CQD films increased mobilities up to 300 $cm^2/Vs$, proving that the contact resistance in the QD-based devices is a major obstacle.[9] To exploit the potential of QDs in practical devices, hybrid systems have been proposed, combining sensitive QDs with high-mobility charge transfer layers. One of the first hybrid detectors uses QD layers deposited on graphene transistors.[10] Photogenerated carriers in QDs are transferred to graphene for fast charge collection, owing to high mobility of exfoliated graphene which is about 1000 $cm^2/Vs$ in that particular device structure. An internal gain mechanism boosts responsivity as charges circulate through the device many times. Therefore, graphene-QD devices reach responsivities as high as $10^7$ A/W at 950 nm. Nevertheless, high dark currents are an issue that leads to high power consumption, as graphene is a semi-metal and current flow through graphene transistors cannot be switched off by electrostatic gating.

As a solution, hybrid photodetectors that use transition metal dichalcogenides (TMDCs) as high mobility transport layers have proven effective in both absorbing infrared radiation in the QD layer and transferring the photogenerated carriers via the TMDC layer to the contacts for charge collection.[11] With mobilities in the 0.5-3 $cm^2/Vs$ range for $MoS_2$ single layers, reaching 200 $cm^2/Vs$ after encapsulation[12], and mobilities of 140 $cm^2/Vs$ [13] for $WS_2$, those two TMDC materials have been reported in hybrid photodetectors.[14–16] PbS-TMDC detectors have been reported with compelling performance up to 1.5 µm, exhibiting detectivities up to $10^{11}$ Jones at the exciton wavelength (around 1 µm)[15]; however, access to longer wavelengths towards 2 µm has been achieved only with the use of HgTe CQDs[17]. In this paper, we employ large PbS CQDs in combination with two kinds of TMDC ($MoS_2$ and $WS_2$) to demonstrate the first hybrid photodetectors with low noise and high detectivity up to 2 µm.

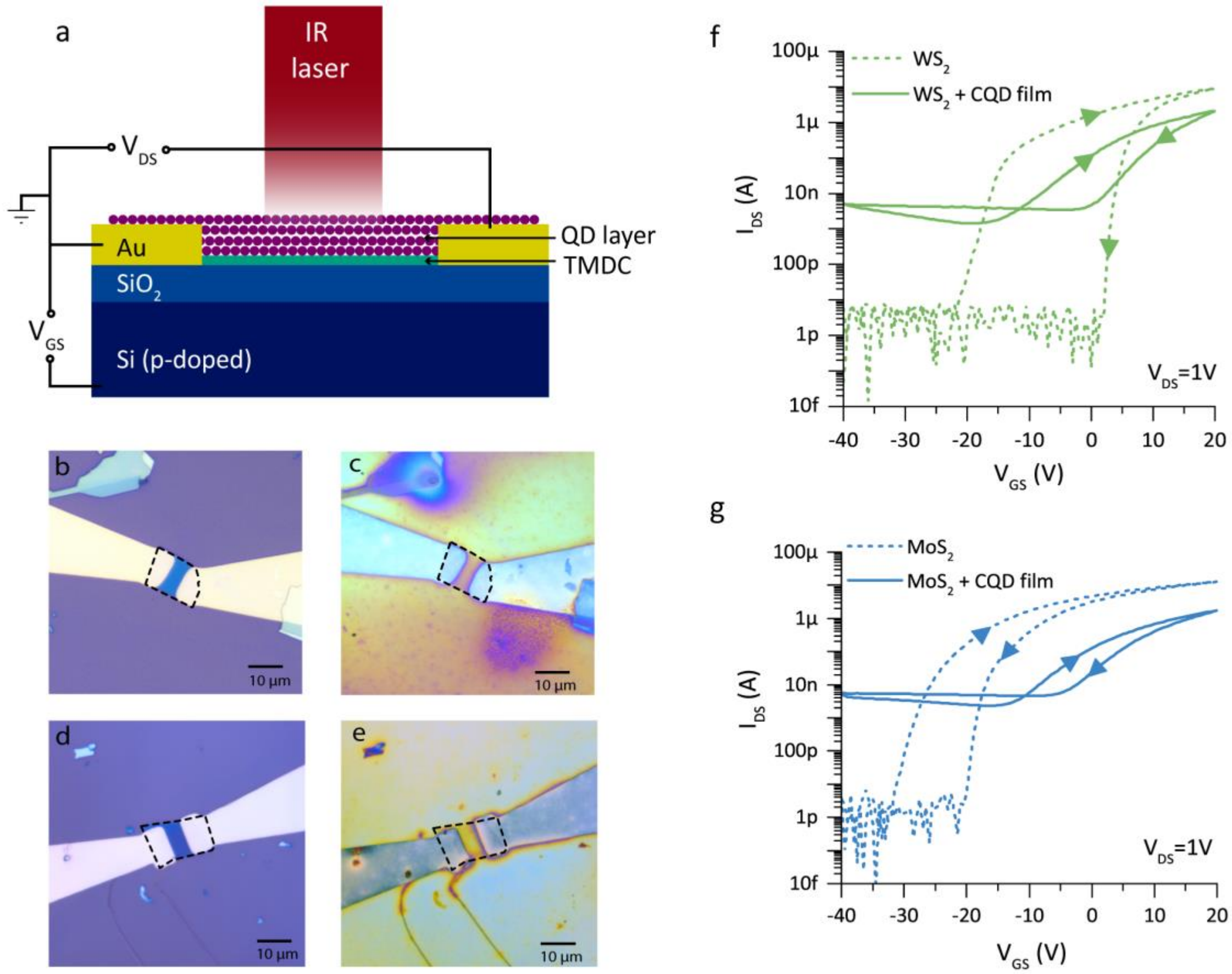

*Figure 1. (a) Hybrid photodetector architecture with circuit diagram. Optical microscope images of few-layer $WS_2$ (b, c) and $MoS_2$ (d, e) transistors before and after PbS CQD deposition, respectively. Areas with TMDCs are indicated by black dotted lines. Transfer curves in dark for $WS_2$ based device (f) and $MoS_2$ based device (g) in different stages of fabrication in logarithmic scale. The arrows correspond to the direction of the gate voltage sweep, indicating the hysteresis.*

Figure 1a depicts the architecture of the hybrid detectors used in this study. The device active area consists of few-layers (5 to 6) of TMDCs – either $MoS_2$ or $WS_2$ – exfoliated on top of a $Si/SiO_2$ substrate. Regions of desired number of thicknesses of TMDCs are identified with optical microscopy and confirmed by Raman spectroscopy (Figure S1). We incorporated few-layers of TMDCs in our devices since previous reports show that FETs which consist of 5-6 layers of TMDCs exhibit lower noise than monolayers or bulk.[18] Gold source and drain contacts are patterned on both edges of TMDCs with photolithography to create TMDC transistors, with an active area of approximately 50 $\mu m^2$.

PbS QDs are synthesized under inert conditions as described in previous work.[15] The details of the synthesis can be found in the methods section. After synthesis, we analyzed the CQDs to determine the exciton wavelength and shape distribution (Figures S2 and S3) to reveal that the exciton absorption peak lies near 1.8 µm, with an average QD diameter of 8.03 ± 1.67 nm. To form the hybrid device and coat the QDs on TMDC transistors as a film, it is necessary to decrease the distance between QDs and to passivate their surfaces with a proper ligand exchange treatment. One of the prominent ligands used in this exchange process is ethanedithiol (EDT).[19] Several devices have been fabricated using EDT, the ligand of choice for previous reports utilizing larger bandgap PbS QDs. However, EDT proved not to be a suitable ligand for the size of PbS QDs considered in this study, leading to poor performance, as explained in detail in Supporting Information (Figure S7). Previous research in our group on PbS solar cells has demonstrated that a mixed ligand treatment of $ZnI_2$ and MPA (zinc iodide and 3-mercaptopropionic acid) serves as a better ligand exchange scheme for large sized PbS QDs and was reported to suppress trap state density, possess higher mobilities (compared to EDT treated QDs) and result in a larger electron affinity.[20] Therefore, this mixed ligand treatment has been our choice for our devices. The CQD solution is spincoated on a layer-by-layer fashion using $ZnI_2$+MPA for ligand exchange to create a QD film on top of the TMDCs, with a thickness of around 85 nm. Optical microscope images before and after QD film deposition can be seen in Figure 1b-c and Figure 1d-e.

We performed measurements in dark conditions using an electromagnetically isolated probe station and a semiconductor analyzer at various stages of fabrication. After the QD layer is deposited on the TMDC transistors, off-state currents increase in both $MoS_2$ and $WS_2$ based devices, as shown in Figure 1f-g. In $MoS_2$ based devices, this can be linked to the presence of sulfur vacancies that create high density of localized states and an uncontrolled doping resulting in an increase of the off-state currents.[21] A similar mechanism can also be present in $WS_2$ based devices. Dark currents in both $WS_2$ and $MoS_2$ based devices are in a similar range in the final devices. From the transfer curves we are able to obtain the mobilities of our devices (see methods section and Table S1). Upon QD deposition, for $WS_2$ based device, mobility decreased from 30.2 to 12.4 $cm^2/Vs$; and for the $MoS_2$ based device, it decreased from 21.9 to 7.3 $cm^2/Vs$. In both devices hysteresis in the transfer curves decreases significantly after the QD layer is deposited. This can be attributed to a decreased interaction of the TMDC layers with adsorbates in the atmosphere.

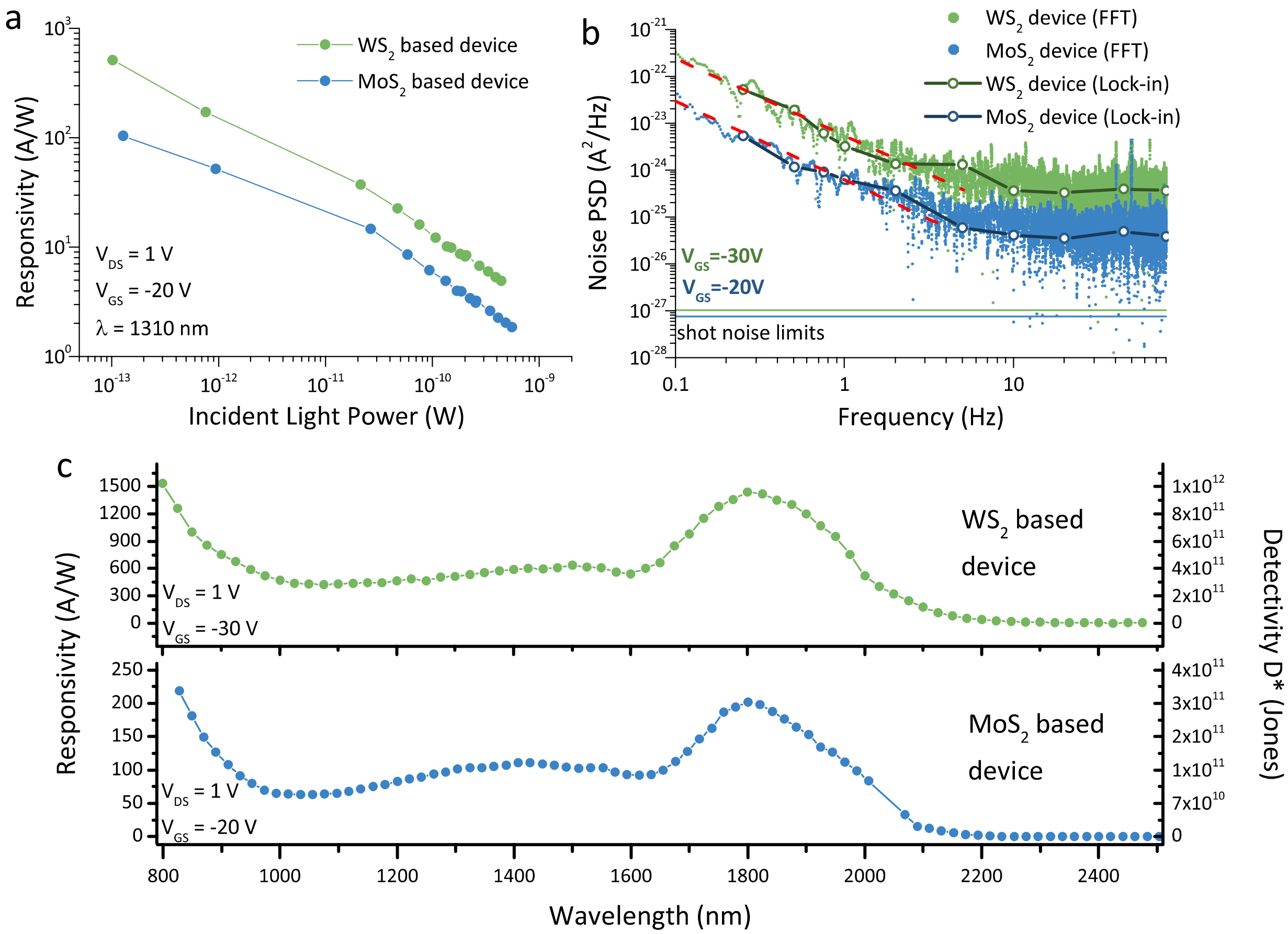

*Figure 2. Optical characterization of our hybrid devices. (a) Responsivity versus incident light power on the photodetector active area, measured by using a 1310 nm laser. (b) Noise characteristics of hybrid devices in dark conditions. Measurements obtained by taking the Fast Fourier Transform (FFT) of the dark current traces (scattered dots) and by the lock-in (hollow dots and lines) match closely. Shot noise limits are indicated by horizontal lines for $WS_2$ (green) and $MoS_2$ (blue) based devices. 1/f behavior (flicker noise) is indicated by dotted red lines as a guide to the eye. (c) Spectral responsivity and detectivity (D*) for the two devices in the 800 – 2400 nm range.*

We then characterized the responsivity of the detectors as a function of light intensity and wavelength. Figure 2a shows a decrease in responsivity with increasing light power, measured using a fiber-coupled 1310 nm laser with adjustable power output. As demonstrated earlier by Kufer et al.[15] hybrid photodetectors become less responsive in high illumination conditions, due to direct charge transfer from QDs. Photogenerated carriers in the QDs induce a reverse electric field, effectively lowering the built-in field.[22] Charge carrier recombination is therefore accelerated at the interface with the resulting drop in responsivity. At 1310 nm, $WS_2$ and $MoS_2$ based devices have a maximum responsivity of 510 A/W and 103 A/W respectively. Photoconductive gain in these detectors originates from photogenerated electrons that are transferred over to the high-mobility TMDC layer. Electrons start to circulate in the circuitry while holes are trapped inside the QDs. This yields to an internal gain, which is determined by the ratio $G=\tau_{lifetime}/\tau_{transit}$, where $\tau_{lifetime}$ is the electron-hole recombination lifetime and $\tau_{transit}$ is the carrier transit time through the contacts.

$\tau_{lifetime}$ values can be extracted from Figure 3a and $\tau_{transit}$ can be calculated as $L^2/\mu V_{DS}$ , where $L$ is the length of the device (5 µm in our devices), µ is the field effect carrier mobility, and $V_{DS}$ is the source-drain voltage (1V for our measurements). We can determine the gain for $WS_2$ based device as approximately $10^7$ and for the $MoS_2$ based device it is about $8.7x10^5$. In order to measure the detectivity (D*) of our detectors we performed noise current measurements in dark. We measured the noise power spectral density (PSD) by two different methods. Firstly, by analyzing the Fourier transform (FFT) of the dark current and secondly, by measuring the noise at different frequencies with a lock-in amplifier, using the same bias voltages as in the responsivity measurements. Noise PSDs for both devices are shown in Figure 2b and they are similar in terms of magnitude and shape. Both FFT and lock-in methods give similar results. In low frequency range, up to approximately 5 Hz, the devices exhibit 1/f behavior (flicker noise, calculated to be $1/f^{1.7}$), indicated by dotted red lines. Figure 2c shows photoresponse in the 0.8-2.2 µm range, with a clear excitonic peak at around 1.8 µm. At this wavelength, responsivity reaches 1442 A/W and 202 A/W for $WS_2$ and $MoS_2$ based devices respectively. The detectivity, D*, is then calculated by:

$$D^* = \frac{R\sqrt{AB}}{i_n} \qquad (1)$$

where $R$ is responsivity, $A$ is device area, $B$ is bandwidth and $i_n$ is the noise current. The spectral D* of the detectors is also plotted in Figure 2c. Our $WS_2$ and $MoS_2$ devices exhibit detectivities of $1.0x10^{12}$ and $2.8x10^{11}$ Jones, respectively, at room temperature with their response reaching up to 2.1 µm. Previously reported PbS-based hybrid photodetectors, while having high detectivites ($7x10^{13}$ Jones), have only achieved photoresponse up to 1.5 µm.[11] The only other hybrid devices to achieve responsivity up to 2.1 µm were based on HgTe, with detectivities in the same order as our $WS_2$ based devices.[17] (see Table S3 for a comparison with previously reported hybrid photodetectors)

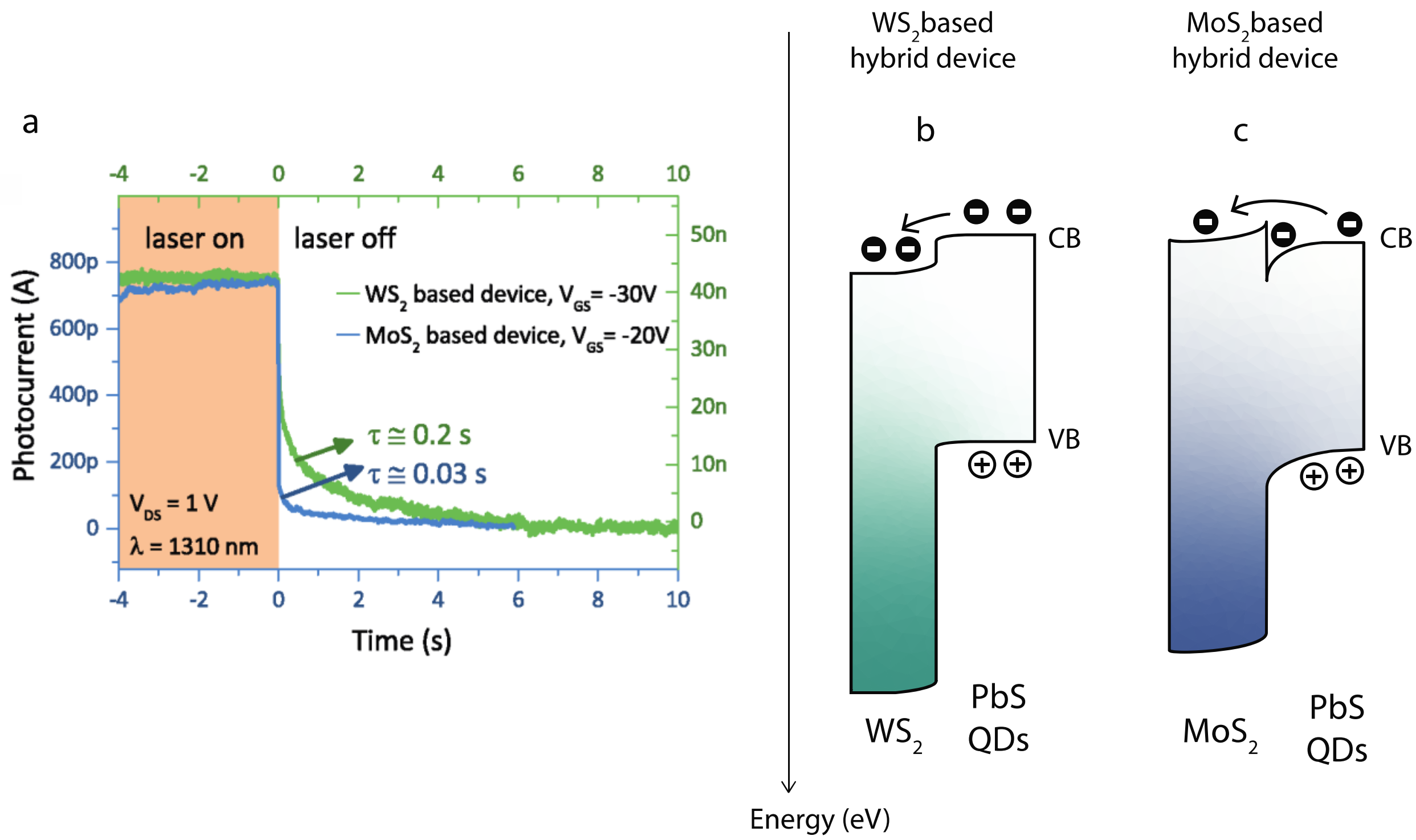

*Figure 3. (a) Photoresponse decay times of the photodetectors, and time constants extracted from the decay curves. Band diagrams of hybrid devices for $WS_2$ (b) and $MoS_2$ (c) based devices. Boxes indicate the bandgap of the materials with the conduction band (CB) and valance band (VB) as their edges.*

Photoresponse speed is another relevant figure of merit of photodetectors. In this sense, $MoS_2$ based devices operate faster than $WS_2$ based ones, as shown in Figure 3a. Upon excitation with a modulated 1310 nm laser, the $MoS_2$ based device has decay response with a time constant of about 0.032 s, while this value is 0.2 s for the $WS_2$ based device. These values lie in the range of previously reported hybrid devices (Table S3). To account for the observed differences in the time response of the $MoS_2$ and $WS_2$ based detectors, we plot in Figure 3b-c the corresponding band alignment of the QD/TMDC heterostructures. We have constructed these band diagrams from ultraviolet photoelectron spectroscopy (UPS) values from literature and our observations of the transfer curves. Further details and discussion can be found in the supplementary information (Figures S5-S6). The clear formation of a type-II heterojunction in the case of $WS_2$ leads to slow recombination between photogenerated electrons transferred to the $WS_2$ and trapped photogenerated holes in the QD layer. In the case of $MoS_2$ based detectors the band alignment creates an electron-rich zone at the interface that facilitates faster recombination, thereby leading to faster photocurrent relaxation. The nature of the band alignment in those two cases may also account for the difference in responsivities between the two types of detectors. $WS_2$ based photodetectors possessing a clear type II junction allows for more efficient electron transfer to the TMDC material compared to the case of the $MoS_2$ based detector. The recorded responsivities of the $MoS_2$ devices have been found to be statistically lower than in the case of $WS_2$ based ones (In Table S2 we present results for the range of values recorded from different devices).

## CONCLUSION

In conclusion, we demonstrated two different types of TMDC-QD hybrid photodetectors, one with $WS_2$ and the other with $MoS_2$, both capable of operating at wavelengths above 1.8 µm. Owing to the semiconducting nature of few-layers of TMDCs, our devices exhibit reduced dark current and noise as compared with graphene-based hybrid photodetectors; therefore, reducing power consumption and increasing detectivity. Favorable band alignments in the $WS_2$ based device gave rise to detectivities of $10^{12}$ Jones at 1.8 µm and room temperature. These hybrid devices have great potential as they are thin and can operate without external cooling, with high sensitivities across the SWIR. Our work also highlights the importance of the selection of the appropriate TMDC channel for a given QD material.

## METHODS

### TMDC transistor fabrication

$Si/SiO_2$ (285 nm) substrates are cleaned in acetone and isopropanol with sonication. $WS_2$ and $MoS_2$ flakes (purchased from 2D semiconductors) are exfoliated by using a PDMS tape on different substrates. After transfer, samples are soaked in acetone for 30 min to remove the residual glue from the tape. Regions of few-layers of TMDCs are identified under an optical microscope. Photoresist, AZ5214E, is spincoated on the samples and baked. Using laser-writing photolithography, source and drain contacts are patterned on the edges of the TMDC flakes. After development, 3 nm of Ti and 50 nm of Au is deposited using an e-beam and thermal evaporator. Following lift-off, samples are annealed under nitrogen atmosphere at 150ºC for 2 hours to improve contact adhesion.

### PbS QD synthesis and deposition

The PbS QDs are synthesized by a previously reported multi-injection method.[23] Briefly, 0.45 g lead oxide (PbO), 100 mL 1-octadecene (ODE) and 3.8 mL oleic acid were mixed and degassed overnight at 90ºC under vacuum. After the degassing, the solution was placed under Ar, the reaction temperature was set to 115ºC and a solution of 55 µL hexamethyldisilathiane ($(TMS)_2S$) in 3 mL ODE was quickly injected. After 6 minutes, a second solution of 135 µL $(TMS)_2S$ in 12 mL ODE was dropwise injected in a rate of 0.75 mL/min. After this second injection, the heating was stopped, and the solution was let to cool down naturally to room temperature. QDs were purified three times by precipitation with acetone and ethanol and redispersing in anhydrous toluene. Finally, the concentration was adjusted to 30 mg/mL and the solution was bubbled with $N_2$.

PbS CQD films are deposited using a layer-by-layer spin-coating process under ambient conditions at 2500 rpm, following the recipe in previous work.[20] $ZnI_2$+MPA (zinc iodide and 3-mercaptopropionic acid) ligand solution is prepared in methanol ($25\times10^{-3}$ M $ZnI_2$ with 0.01% MPA v/v). Dropping CQD solution (in toluene) following by ligand solution and methanol for washing the excess unbound ligands, layer-by-layer deposition is performed on the TMDC transistors. For the devices with ethanedithiol (EDT), a ligand solution of (II) 1,2-Ethanedithiol (EDT) in acetonitrile (ACN) (0.01% v/v) is used. In this case, the washing chemicals are replaced with ACN and toluene. Using this layer-by-layer spincasting technique, 4 layers are deposited in the substrates, resulting in a thickness of approximately 85 nm of QD film on our TMDC transistors.

**Electrical and Optical Characterization**

Transfer curves before and after QD deposition, as well as noise measurements are performed in an electromagnetically isolated probe station with a semiconductor analyzer (Agilent B1500A) using a two-probe connection in ambient. Noise measurements are taken as Fourier transform of dark current traces spaced 4 ms apart and are further confirmed by a lock-in based setup (Zurich Instruments MFLI with a 2-channel low noise power source Keysight B2962A) in the same probe station.

By fitting the linear region of the transistor transfer curves, mobility, $\mu$, is calculated under the short channel approximation as:

$$\mu = \frac{dI_{DS}}{dV_{GS}} \frac{L}{W C_{ox} V_{DS}} \tag{2}$$

where $I_{DS}$ is the source-drain current, $V_{GS}$ is the back-gate voltage, $L$ and $W$ are the length and width of the TMDC channel respectively; $C_{ox}$ is the capacitance of the oxide layer and $V_{DS}$ is the source-drain voltage (set as 1 V for our measurements). $C_{ox}$ is taken as $1.28\times10^{-8}$ F $cm^{-2}$. The slopes of the transfer curves are calculated when the device is in the on (conducting, non-depleted) state. A significant hysteresis in the transfer characteristics can be seen in all devices, and can be linked to atmospheric adsorption and trapped charges; an issue being addressed by many groups.[24,25] Due to hysteresis, forward and backward mobilities are different and are tabulated and further explained in the Table S1.

Light power dependent responsivity measurements are performed in the same probe station using a fiber-coupled 1310 nm laser with light power controlled by an Agilent A33220A waveform generator. Spectral responsivity measurements are performed by a fiber coupled supercontinuum laser (SuperKExtreme EXW-4, NKT Photonics). We performed measurements at gate voltages where the photoresponse was highest. Laser power density measurements are performed by a calibrated Newport power meter with Si and Ge detectors using a 25 µm diameter circular aperture right before the detector to account for the beam divergence in different wavelengths.

## ACKNOWLEDGEMENTS

The research leading to these results has received funding from the H2020 Programme under grant agreement no. 785219 Graphene Flagship. The authors also acknowledge financial support from the European Research Council (ERC) under the European Union's Horizon 2020 research and innovation programme (grant agreement no. 725165), the Spanish Ministry of Economy and Competitiveness (MINECO), and the "Fondo Europeo de Desarrollo Regional" (FEDER) through grant TEC2017-88655-R. The authors also acknowledge financial support from Fundacio Privada Cellex, the program CERCA and from the Spanish Ministry of Economy and Competitiveness, through the "Severo Ochoa" Programme for Centres of Excellence in R&D (SEV-2015-0522). I. Ramiro acknowledges support from the Ministerio de Economía, Industria y Competitividad of Spain via a Juan de la Cierva fellowship (FJCI-2015-27192).

**SUPPORTING INFORMATION**

Supporting Information includes Raman measurements of TMDC layers, spectral absorption measurements of PbS CQDs used in the devices as well as TEM images, test device characterization without TMDC layers, discussions on mobility and hysteresis, observations about transfer curves, discussion about constructing the band diagrams leading to Figure 3b-c, characterization of the hybrid devices with EDT ligand exchange, statistics of several hybrid devices and comparison with the state of the art hybrid photodetectors.

## REFERENCES

(1) Rogalski, A. Recent Progress in Infrared Detector Technologies. *Infrared Phys. Technol.* **2011**, *54* (3), 136–154.

(2) Jiang, L. J.; Ng, E. Y. K.; Yeo, A. C. B.; Wu, S.; Pan, F.; Yau, W. Y.; Chen, J. H.; Yang, Y. A Perspective on Medical Infrared Imaging. *J. Med. Eng. Technol.* **2005**, *29* (6), 257–267.

(3) Mikhailova, M.; Stoyanov, N.; Andreev, I.; Zhurtanov, B.; Kizhaev, S.; Kunitsyna, E.; Salikhov, K.; Yakovlev, Y. Optoelectronic Sensors on GaSb- and InAs- Based Heterostructures for Ecological Monitoring and Medical Diagnostics. In *Proc.SPIE*; 2007; Vol. 6585.

(4) Rogalski, A. *Infrared Detectors*, Second Edition; CRC Press, 2011.

(5) Talapin, D. V.; Lee, J.-S.; Kovalenko, M. V.; Shevchenko, E. V. Prospects of Colloidal Nanocrystals for Electronic and Optoelectronic Applications. *Chem. Rev.* **2010**, *110* (1), 389–458.

(6) Moreels, I.; Lambert, K.; Muynck, D. De; Vanhaecke, F.; Poelman, D.; Martins, J. C.; Allan, G.; Hens, Z. Size-Dependent Optical Properties of Colloidal PbS Quantum Dots. *ACS Nano* **2009**, *3* (10), 3023–3030.

(7) Kagan, C. R.; Murray, C. B. Charge Transport in Strongly Coupled Quantum Dot Solids. *Nat. Nanotechnol.* **2015**, *10* (12), 1013–1026.

(8) Balazs, D. M.; Bijlsma, K. I.; Fang, H. H.; Dirin, D. N.; Döbeli, M.; Kovalenko, M. V.; Loi, M. A. Stoichiometric Control of the Density of States in PbS Colloidal Quantum Dot Solids. *Sci. Adv.* **2017**, *3* (9), 1–8.

(9) Dolzhnikov, D. S.; Zhang, H.; Jang, J.; Son, J. S.; Panthani, M. G.; Shibata, T.; Chattopadhyay, S.; Talapin, D. V. Composition-Matched Molecular “Solders” for Semiconductors. *Science (80-. ).* **2015**, *347* (6220), 425–428.

(10) Konstantatos, G.; Badioli, M.; Gaudreau, L.; Osmond, J.; Bernechea, M.; de Arquer, F. P. G.; Gatti, F.; Koppens, F. H. L. Hybrid Graphene–Quantum Dot Phototransistors with Ultrahigh Gain. *Nat. Nanotechnol.* **2012**, *7* (6), 363–368.

(11) Hu, C.; Dong, D.; Yang, X.; Qiao, K.; Yang, D.; Deng, H.; Yuan, S.; Khan, J.; Lan, Y.; Song, H.; Tang, J. Synergistic Effect of Hybrid PbS Quantum Dots/2D-$WSe_2$ Toward High Performance and Broadband Phototransistors. *Adv. Funct. Mater.* **2017**, *27* (2).

(12) Félix, G.; Thomas, N. Single-Layer $MoS_2$ Transistors. *Phys. Rev. E - Stat. Physics, Plasmas, Fluids, Relat. Interdiscip. Top.* **2004**, *70* (5), 16.

(13) Ovchinnikov, D.; Allain, A.; Huang, Y.-S.; Dumcenco, D.; Kis, A. Electrical Transport Properties of Single-Layer $WS_2$. *ACS Nano* **2014**, *8* (8), 8174–8181.

(14) Kufer, D.; Lasanta, T.; Bernechea, M.; Koppens, F. H. L.; Konstantatos, G. Interface Engineering in Hybrid Quantum Dot-2D Phototransistors. *ACS Photonics* **2016**, *3* (7), 1324–1330.

(15) Kufer, D.; Nikitskiy, I.; Lasanta, T.; Navickaite, G.; Koppens, F. H. L.; Konstantatos, G. Hybrid 2D-0D $MoS_2$ -PbS Quantum Dot Photodetectors. *Adv. Mater.* **2015**, *27* (1), 176–180.

(16) Yu, Y.; Zhang, Y.; Song, X.; Zhang, H.; Cao, M.; Che, Y.; Dai, H.; Yang, J.; Zhang, H.; Yao, J. PbS-Decorated $WS_2$ Phototransistors with Fast Response. *ACS Photonics* **2017**, *4* (4), 950–956.

(17) Huo, N.; Gupta, S.; Konstantatos, G. $MoS_2$ -HgTe Quantum Dot Hybrid Photodetectors beyond 2 µm. *Adv. Mater.* **2017**, *29* (17), 1606576.

(18) Sharma, D.; Motayed, A.; Shah, P. B.; Amani, M.; Georgieva, M.; Glen Birdwell, A.; Dubey, M.; Li, Q.; Davydov, A. V. Transfer Characteristics and Low-Frequency Noise in Single- and Multi-Layer $MoS_2$ Field-Effect Transistors. *Appl. Phys. Lett.* **2015**, *107* (16), 1–6.

(19) Klem, E. J. D.; Shukla, H.; Hinds, S.; MacNeil, D. D.; Levina, L.; Sargent, E. H. Impact of

Dithiol Treatment and Air Annealing on the Conductivity, Mobility, and Hole Density in PbS Colloidal Quantum Dot Solids. *Appl. Phys. Lett.* **2008**, *92* (21), 1–4.
(20) Pradhan, S.; Stavrinadis, A.; Gupta, S.; Bi, Y.; Di Stasio, F.; Konstantatos, G. Trap-State Suppression and Improved Charge Transport in PbS Quantum Dot Solar Cells with Synergistic Mixed-Ligand Treatments. *Small* **2017**, *13* (21), 1700598.
(21) Sim, D. M.; Kim, M.; Yim, S.; Choi, M. J.; Choi, J.; Yoo, S.; Jung, Y. S. Controlled Doping of Vacancy-Containing Few-Layer $MoS_2$ via Highly Stable Thiol-Based Molecular Chemisorption. *ACS Nano* **2015**, *9* (12), 12115–12123.
(22) Li, M.; Chen, J. S.; Routh, P. K.; Zahl, P.; Nam, C. Y.; Cotlet, M. Distinct Optoelectronic Signatures for Charge Transfer and Energy Transfer in Quantum Dot–$MoS_2$ Hybrid Photodetectors Revealed by Photocurrent Imaging Microscopy. *Adv. Funct. Mater.* **2018**, *28* (29), 1–9.
(23) Lee, J. W.; Kim, D. Y.; Baek, S.; Yu, H.; So, F. Photodetectors: Inorganic UV-Visible-SWIR Broadband Photodetector Based on Monodisperse PbS Nanocrystals. *Small* **2016**, *12* (10), 1246–1246.
(24) Liu, S.; Yuan, K.; Xu, X.; Yin, R.; Lin, D. Y.; Li, Y.; Watanabe, K.; Taniguchi, T.; Meng, Y.; Dai, L.; Ye, Y. Hysteresis-Free Hexagonal Boron Nitride Encapsulated 2D Semiconductor Transistors, NMOS and CMOS Inverters. *Adv. Electron. Mater.* **2019**, *5* (2), 2–7.
(25) Wang, J.; Rhodes, D.; Feng, S.; Nguyen, M. A. T.; Watanabe, K.; Taniguchi, T.; Mallouk, T. E.; Terrones, M.; Balicas, L.; Zhu, J. Gate-Modulated Conductance of Few-Layer $WSe_2$ Field-Effect Transistors in the Subgap Regime: Schottky Barrier Transistor and Subgap Impurity States. *Appl. Phys. Lett.* **2015**, *106* (15).

**For Table of Contents Use Only**

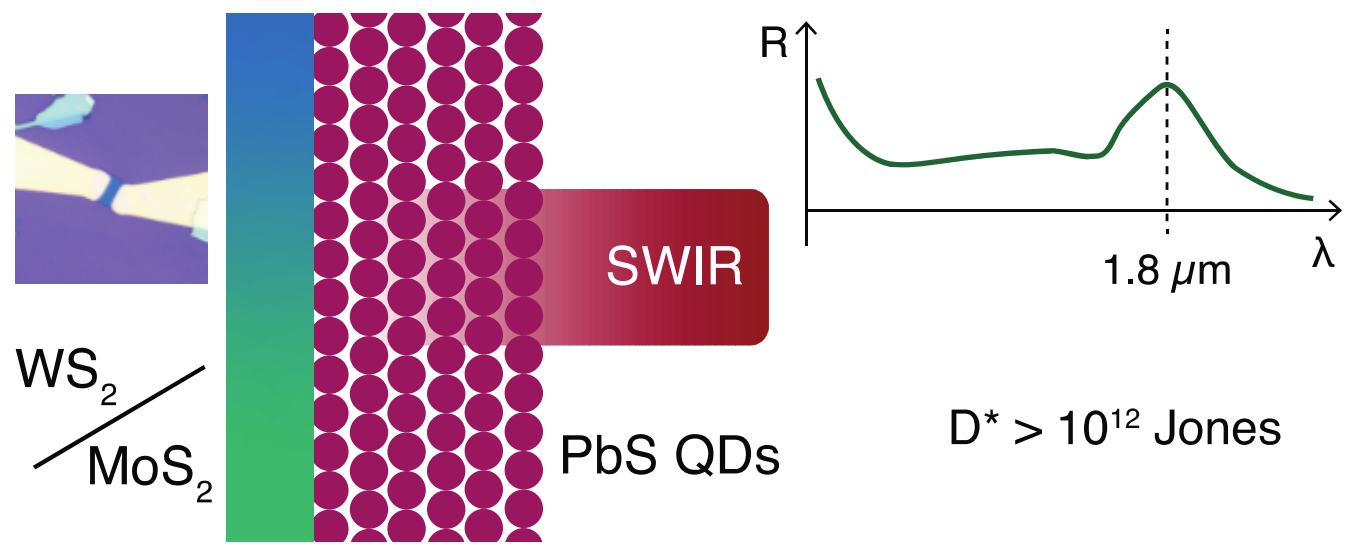

**Manuscript Title:** High Sensitivity Hybrid PbS CQD - TMDC Photodetectors up to 2 μm

**Authors:** Onur Özdemir, Iñigo Ramiro, Shuchi Gupta and Gerasimos Konstantatos

**Synopsis:** Table of contents graphic includes an optical microscope image of our $WS_2$ transistor after it is fabricated, a drawing of the device structure that we have employed in the study (blue/green area representing $WS_2$ or $MoS_2$ layers, purple circles representing the PbS QDs with short-wave infrared radiation incident on the device from the right). The image also includes the striking result of the responsivity range covered by our photodetectors, with detectivities exceeding $10^{12}$ Jones.

# Supporting Information

# High Sensitivity Hybrid PbS CQD - TMDC Photodetectors up to 2 μm

Onur Özdemir,† Iñigo Ramiro,† Shuchi Gupta† and Gerasimos Konstantatos*,†,‡

†ICFO – Institut de Ciències Fotòniques, Av. Carl Friedrich Gauss 3, 08860 Castelldefels (Barcelona), Spain
‡ICREA – Institucio Catalana de Recerca i Estudis Avançats, Passeig Lluís Companys 23, 08010 Barcelona, Spain

*This supporting Information includes 11 pages, 7 figures, 3 tables*

## Raman Measurements of exfoliated TMDC layers

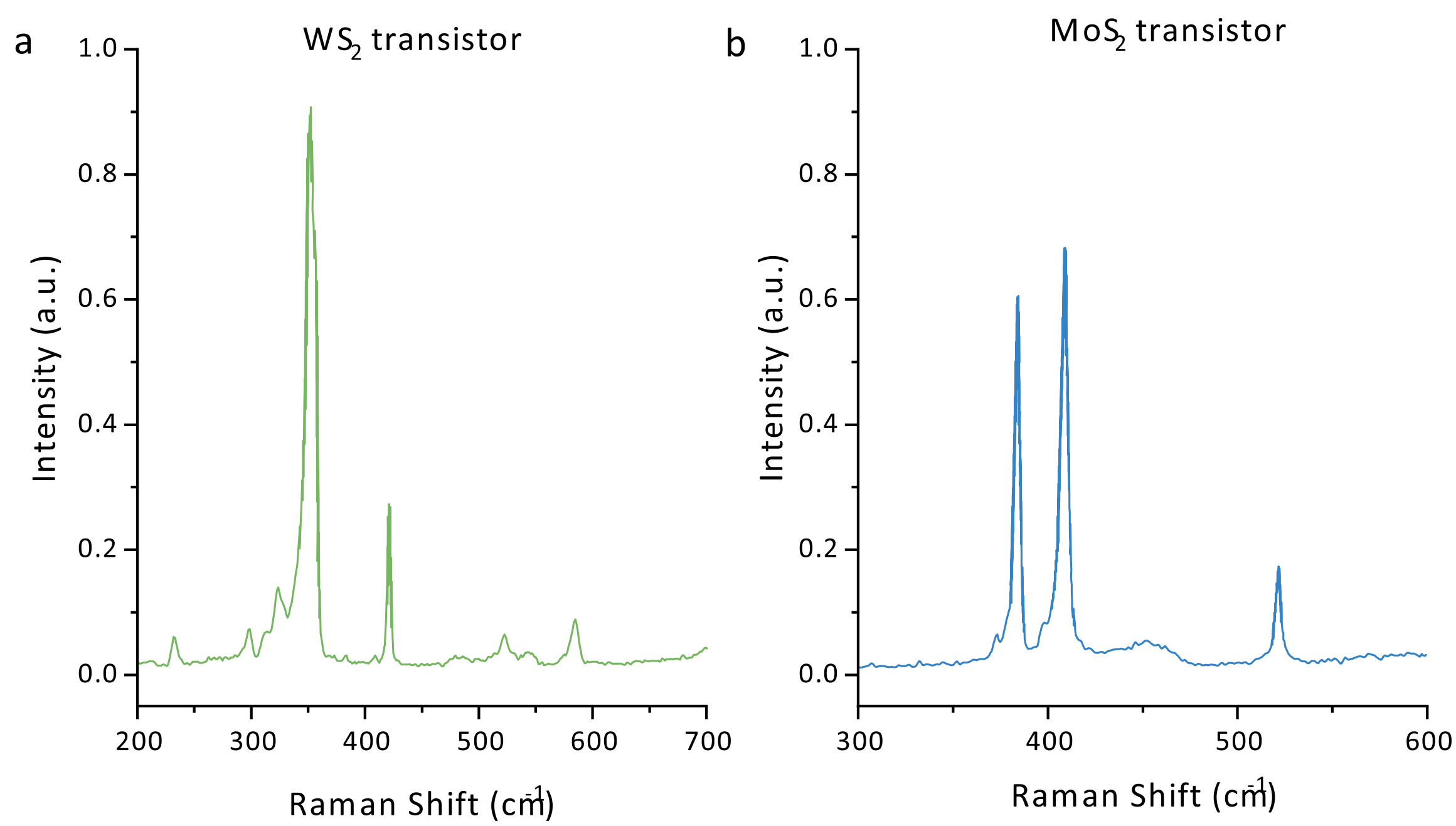

Figure S1. Raman spectra for $WS_2$ (a) and $MoS_2$ (b) after exfoliation

After exfoliation of TMDC layers using the Scotch tape method, we identified regions of few atomic layers to be made into devices and measured their Raman spectra. According to the Raman measurements, $MoS_2$ consists of 5-6 layers and $WS_2$ consists of 3-5 layers.

## Properties of Colloidal PbS Quantum Dots

To investigate the colloidal quantum dots that are synthesized as described in the methods part of the main text, we performed spectrophotometer absorption measurements to determine the spectral response; and TEM (transmission electron microscopy) measurements to determine the size distribution of the dots.

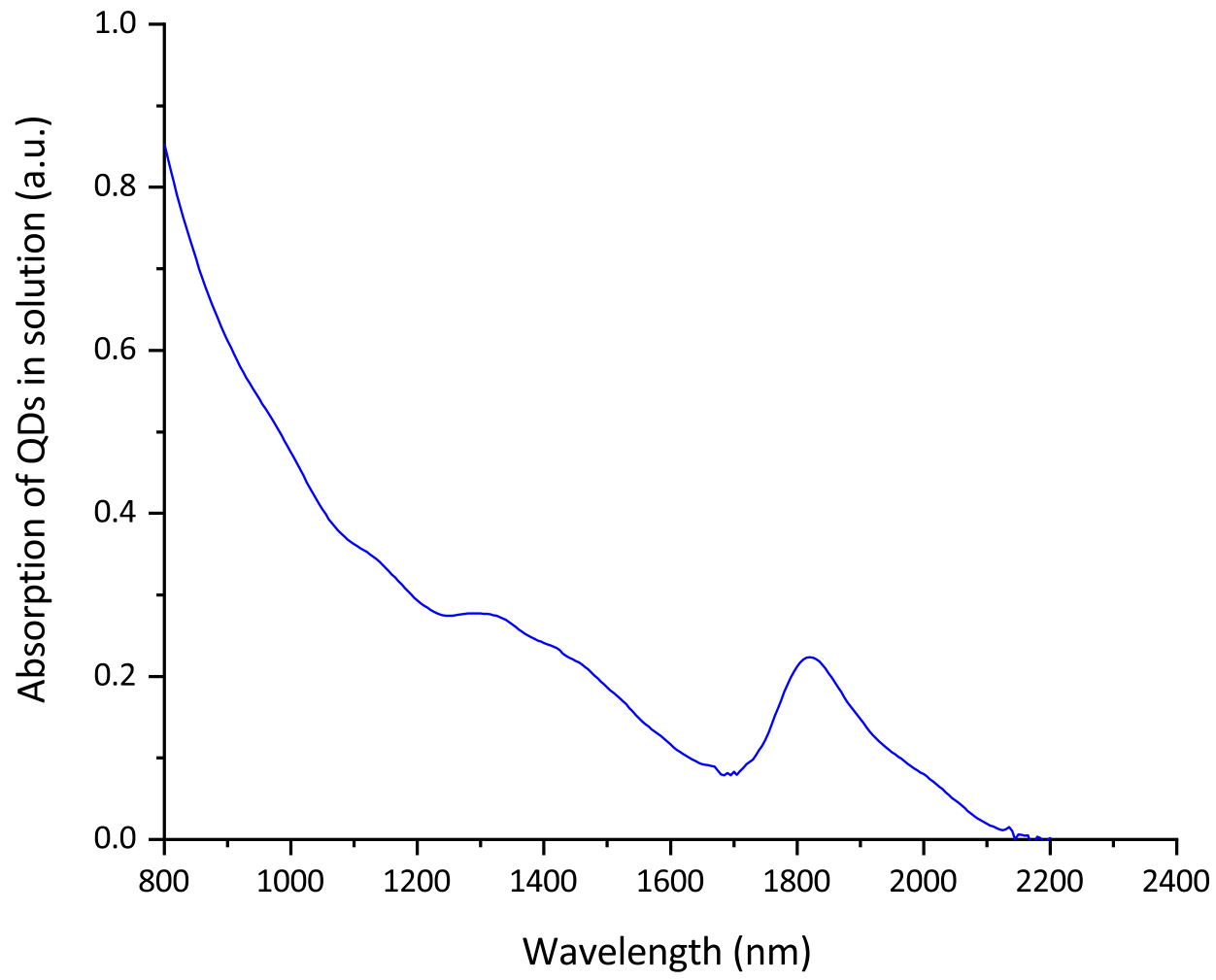

Figure S2. Spectral absorption of the PbS CQDs in toluene after synthesis measured using Cary-5000 spectrophotometer.

Right after the synthesis of PbS colloidal quantum dots, we measured their absorption spectra while in solution of toluene, using a Cary 5000 Spectrophotometer. Pure toluene is taken as a background absorber and QD solution is diluted in the same toluene for absorption measurements. The resulting curve indicates that our dots have an exciton wavelength near 1.8 μm.

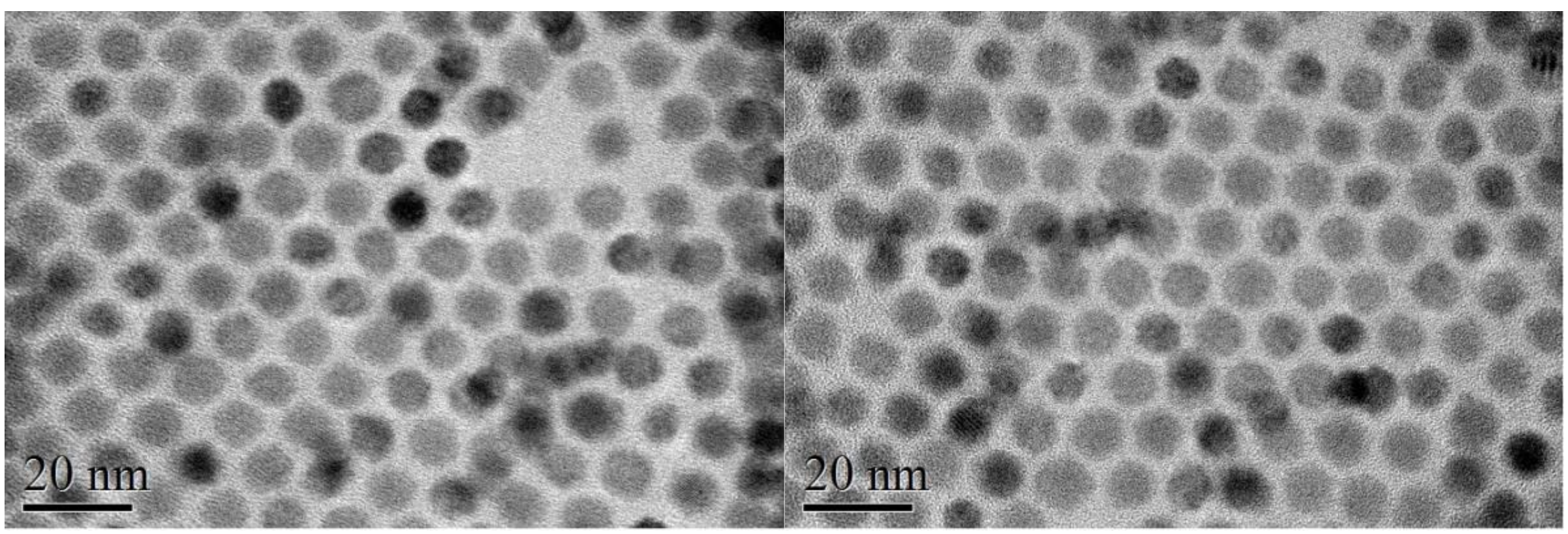

Figure S3. TEM images of the CQDs used in this study. The scale bar on the bottom left indicates 20 nm.

Analysis of TEM images reveal that the average diameter of the dots are 8.03 ± 1.67 nm. These results match closely with the model developed by Moreels et al. [1] and also from previous experimental works in our group.[2]

**Photodetector with only QD layer**

To see the impact of TMDC layer on our QD devices, we fabricated control devices without any TMDC layer on them. CQDs are spincast using the same ligand exchange methods and the resulting transistor is measured in the same conditions as the hybrid devices.

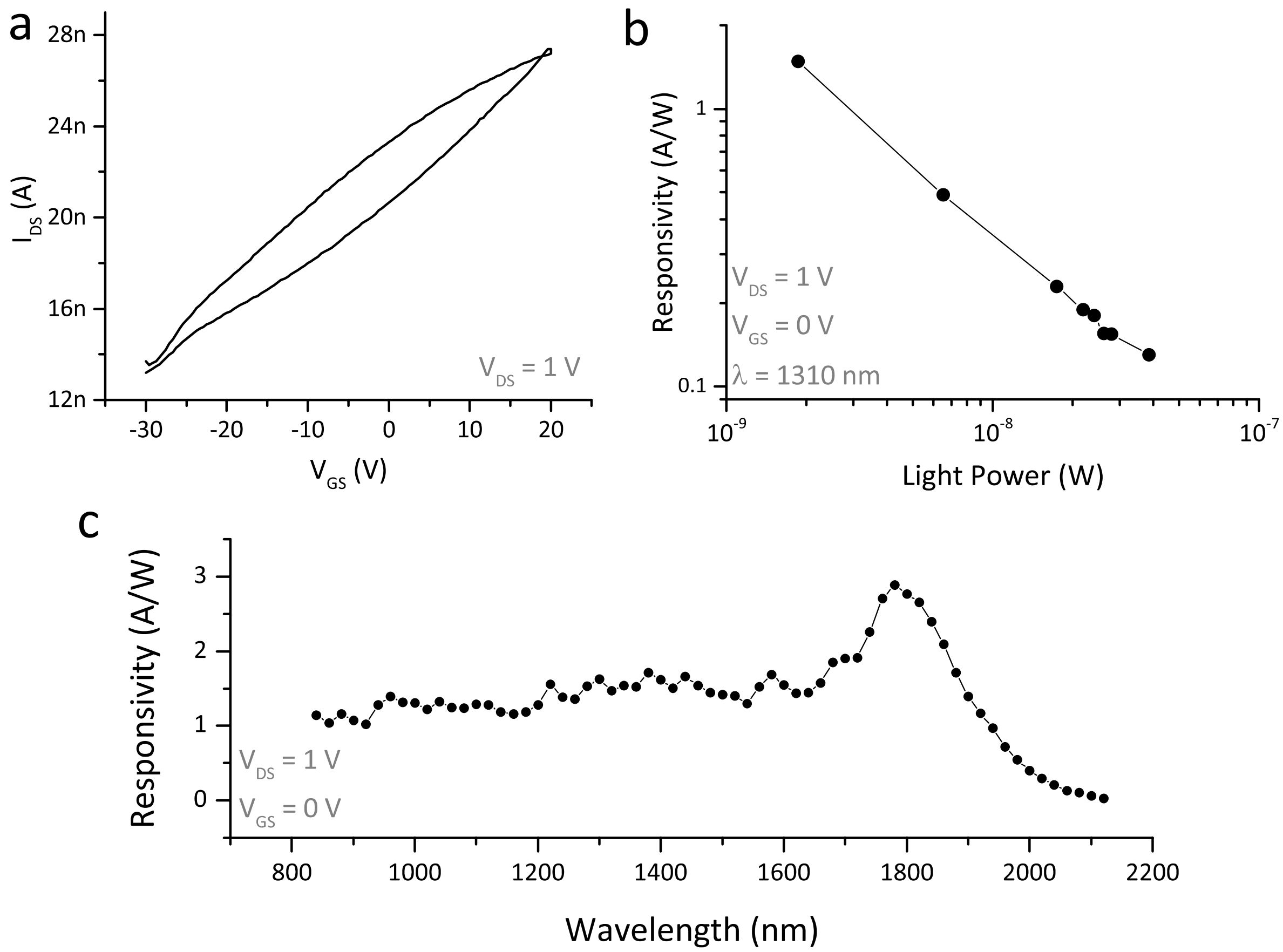

Figure S4. Electrical and optical characterization of QD photodetectors. (a) Transfer curve of the device, showing a hysteresis behavior. (b) Responsivity versus incident light power on the device with a laser of 1310 nm wavelength in logarithmic scale. (c) Responsivity versus wavelength obtained by using a supercontinuum laser.

Figure S4a shows the transfer curve of the QD photodetector. The material is n-type with a calculated mobility of $4.8x10^{-2}$ $cm^2/Vs$. Figure S4b shows the response of this photodetector to 1310 nm laser light with varying intensities. Just as in the hybrid devices, there is a decrease of responsivity as the laser power increases. The minimum measurable signal gives a responsivity of about 1.46 A/W. This is at least 2 orders of magnitude lower than the hybrid devices presented in the main text. Photogenerated carriers recombine before they can be extracted by the contacts due to a combination of low mobility and large contact resistance.

Spectral responsivity is measured using a supercontinuum laser with normalized light intensities as in the main text. The resulting spectral shape is almost identical to the one of hybrid devices, except the increasing responsivity below 900 nm, which can be attributed to the onset of absorption of the TMDCs.

**<u>Mobility Calculations oh Hybrid Devices</u>**

The mobilities of the active layers are calculated using equation (2) in the method section of the main text. Due to hysteresis, two different mobility values are obtained for backward and forward curves of the transfer curve. Forward curve is defined as starting with a negative $V_{GS}$ and increasing to positive voltages, while backward curve is the opposite. In the main text, only the backward mobilities are reported.

| Device | Forward Mobility ($cm^2/Vs$) | Backward Mobility ($cm^2/Vs$) |
|---|---|---|
| $MoS_2$ | 17.74 | 21.93 |
| $MoS_2$ + QDs | 5.10 | 7.27 |
| $WS_2$ | 19.38 | 30.16 |
| $WS_2$ + QDs | 6.19 | 12.42 |

Table S1. Mobilities of the devices in various steps of fabrication. Forward and backward values are extracted from Figure 1f-g in the main text and calculated using Equation (1)

**Transfer Curve Shifts in Hybrid Devices**

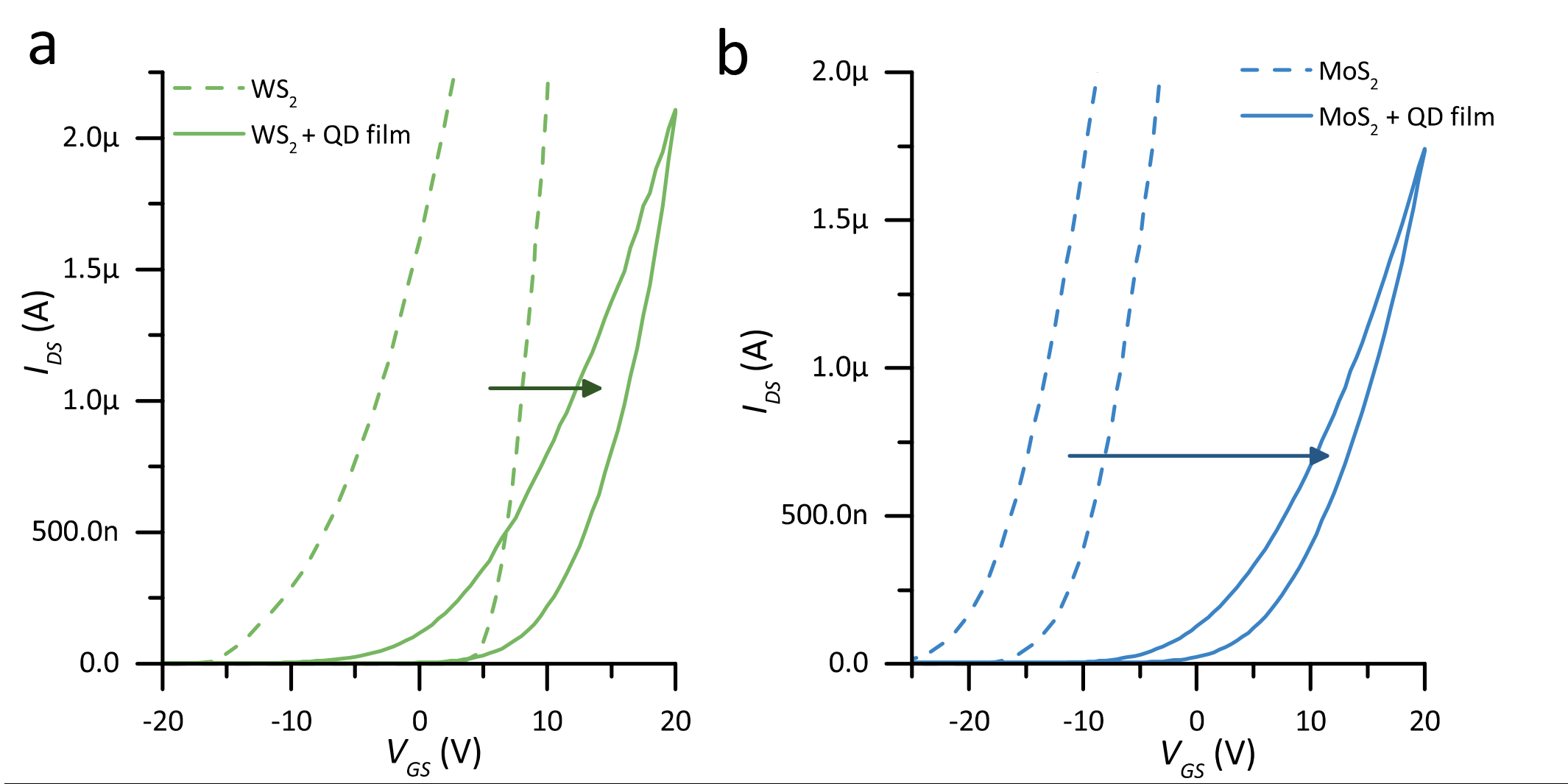

Figure S5. Transfer ($I_{DS}$-$V_{GS}$) curves of the hybrid photodetectors in linear scale for (a) $WS_2$ and (b) $MoS_2$ based devices, including the hysteresis effect. In both cases, curves shift to the right after QD deposition, which is indicated by an arrow.

While investigating the effect of QDs on the TMDC transistor system, looking at the shifts in transfer curves gives us an insight. Figure S5 shows the transfer characteristics of the devices in the main text (Figure 1f-g) in linear scale so that the reader can see the threshold voltages clearly. Threshold voltage is the voltage above which the transistor starts conducting. In both cases, before QD deposition we can see that the threshold voltages lie in the lower voltages. We also observe that the transistors are n-type. After QD deposition, threshold voltages shift to higher voltages. This indicates that the hybrid transistor system is less n-doped than the bare TMDC transistor system, which suggest electron transfer from the TMDC to the QDs upon junction of the two materials. Therefore, we conclude that the fermi level of the QDs must be below the fermi levels of TMDCs before they are combined. The horizontal shift in the transfer curves is larger in the case of $MoS_2$ when compared to $WS_2$. This suggests there is a larger difference in fermi levels between $MoS_2$ layers and QD layers. This difference is lower in $WS_2$. This observation complements our discussion for Figure 3b-c in the main text and Figure S6.

**Band Diagrams of Hybrid Devices**

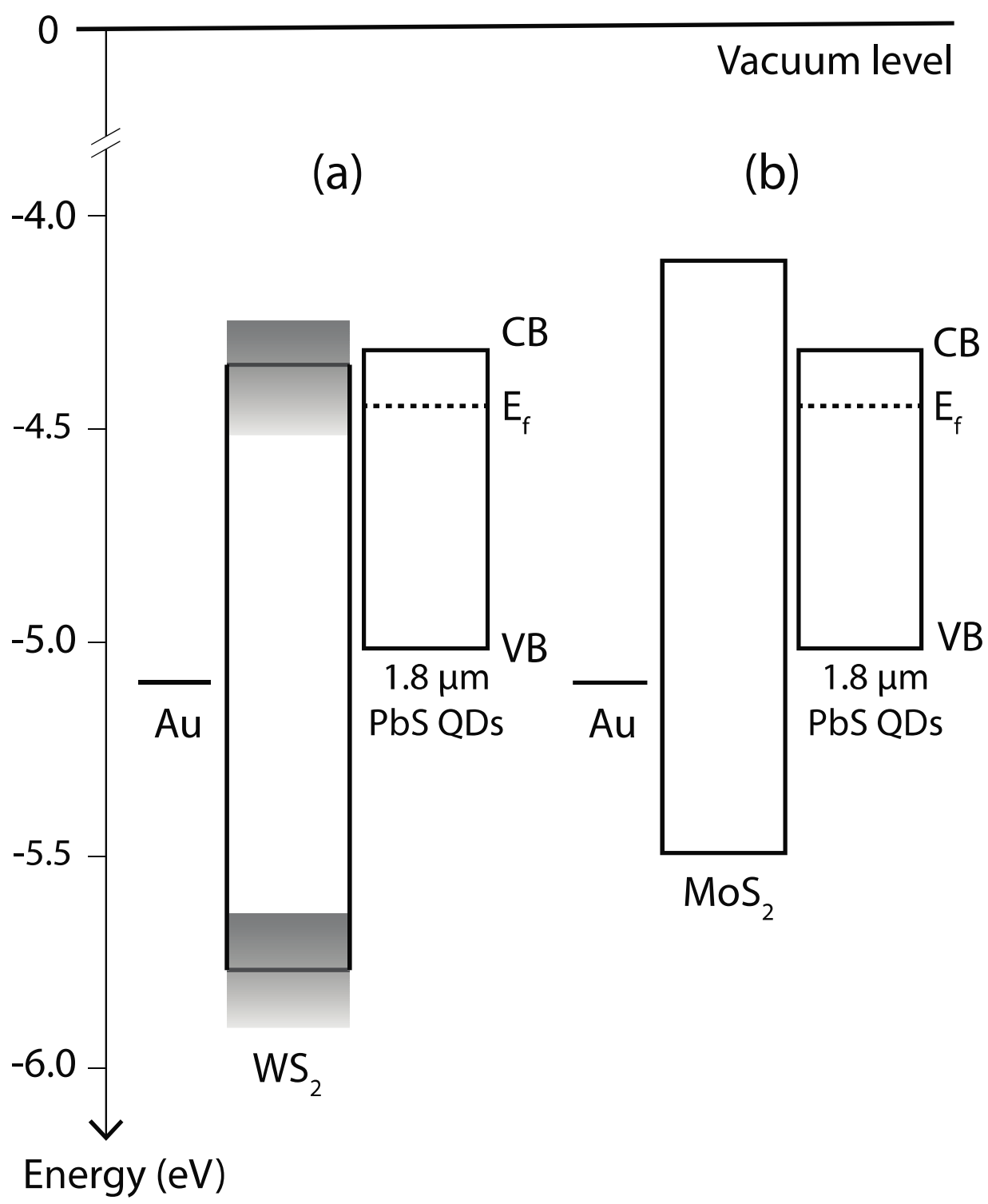

Figure S6. Band Alignments of the materials used in our devices for (a) $WS_2$ based and (b) $MoS_2$ based devices. CB and VB indicate conduction and valance bands respectively. $E_f$ indicates the Fermi level denoted by dashed lines. The shaded gray areas in (a) indicates the uncertainty of the exact position of the band edges for $WS_2$.

To construct the band diagrams of the TMDCs, we have used UPS values from the literature. In this work we have used exfoliated TMDC layers therefore, the sizes of our flakes are usually in the order of 30-100 µm$^2$. In these small sizes, UPS measurements were not possible. Also, for our hybrid devices, PbS QD layer thickness exceeds the penetration depth of UPS. As a result, a detailed analysis of the interface between TMDC-QDs were not possible.

Reports of experimental UPS measurements of exfoliated TMDC flakes show that valence band of $WS_2$ and $MoS_2$ lies below 5.99 and 5.49 eV from the vacuum level respectively.[3] Some other UPS measurements performed on chemically grown $WS_2$ denote the valance band edge around -5.65 eV below the vacuum level. [4] For this reason, in Figure S6, we included and uncertainty in the valance and conduction band edges for $WS_2$, as the exact position is still not settled yet. For the bandgaps of few-layers of TMDCs, we have used the experimental values from other groups; for $MoS_2$ and $WS_2$ they are approximately 1.39 eV [5] and 1.42 eV [6], respectively

As for the PbS QDs, we have used -5.0 eV for the valence band maximum (VBM), as obtained from previous UPS measurements from our group for QDs similar to the ones employed here.[2] The conduction band maximum (CBM) has been obtained by adding the measured optical bandgap to the VBM.

Regarding the position of the Fermi level in each material prior to junction, we have used based our analysis in the previous discussion on the transfer curves of our photodetectors (Figure S5). We know that: (i) the Fermi level of the QDs must be lower than that of the TMDCs, (ii) the Fermi level of $MoS_2$ is higher than that of $WS_s$.; and (iii) both $WS_2$ and $MoS_2$ are n-type and highly conductive, therefore, for each of them, we assume that the Fermi level should not be more than 3 or 4 times *kT* below the CBM. In addition, we have calculated the Fermi level of our PbS CQD film to be approximately 0.085 eV below the CBM, as we explain next.

From the transfer curves of devices containing only PbS QDs (Figure S4), we have been able to obtain the electron population, *n*, in our PbS CQD film, following the method described in Ref. 7. We have obtained n ~ $6.7 \times 10^{17}$ $cm^{-3}$. We have calculated density of states of the CBM, $N_C$, in our QDs considering: (i) the volume occupied by a single QD (a sphere of diameter 8.0 nm); (ii) a QD packing density in the film of 0.64 (for a random distribution); (iii) 8-fold degeneracy of the CBM of PbS QDs.[8] With those values, we get $N_C = 1.9 \times 10^{19}$ $cm^{-3}$. The occupancy factor, *f*, of the CBM is then approximately $f = {}^{n}/_{N_C} = 0.035$.

From Fermi-Dirac statistics, we know:

$$f^{-1} = e^{(E-E_F)/kT} + 1$$

Where *f* is the occupancy of an electronic state with energy *E*, $E_F$ is the Fermi level, *k* is Boltzmann constant and *T* is temperature. Using the value *f* = 0.035 yields to $E - E_F \sim 0.085$ eV.

**TMDC-PbS QD hybrid photodetectors with EDT (ethanedithiol) ligand treatment**

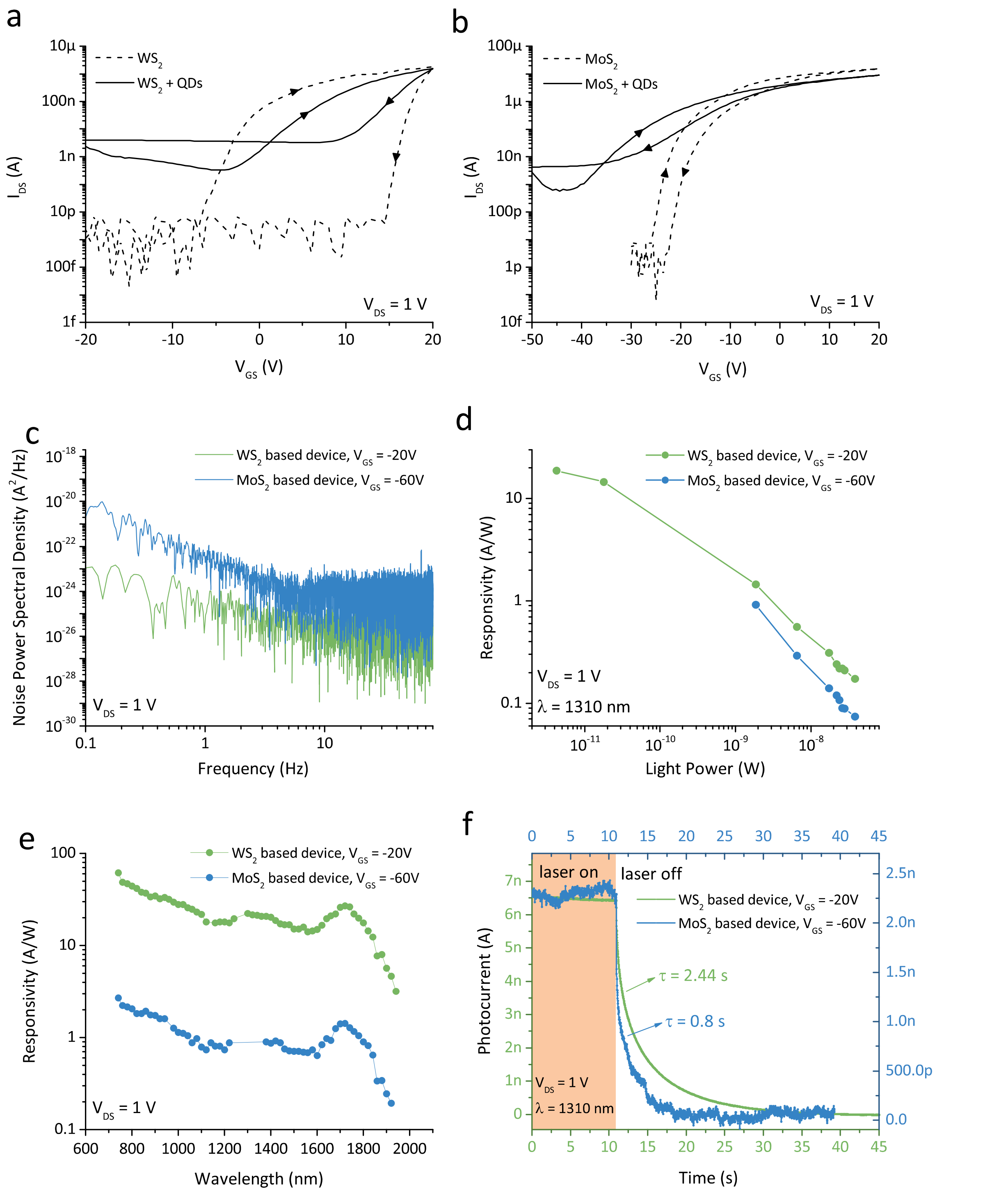

Figure S7. Characterization results for TMDC – PbS QD devices with EDT (ethanedithiol) treatment. Transfer curves for the photodetectors after and before QD deposition for $MoS_2$ (a) and $WS_2$ (b) as carrier channel. Arrows indicate the direction of the voltage scan and highlights

the hysteresis effect. (c) Noise power spectral density measurements obtained by the FFT of dark current traces. (d) Responsivity versus incident light power on the devices with 1310 nm laser wavelength. (e) Spectral responsivities for hybrid devices measured with a supercontinuum laser. (f) Response speed of the hybrid devices to a 1310 nm laser and calculated time constants for each case.

To test the effect of the ligands, hybrid photodetectors are fabricated with $WS_2$ and $MoS_2$ (5 layers) and PbS QDs (1.75 µm exciton wavelength) treated with EDT (ethanedithiol) during spin coating. The same architecture and dimensions are used described in the main text.

It has been previously shown that EDT induces a p-doping rather than an n-doping as in the case of iodine-based ligands, such as $ZnI_2$+MPA.[9] The referenced publication also shows that treating PbS QDs with EDT modifies the conduction and valance bands such that their energies decrease. i.e. they move closer to vacuum level. Following our discussion from before, by observing the transfer curves in Figure S7b, we can say that the threshold voltage after the QD deposition shifts to lower voltages. This indicates that QD deposition induced an n-doping on the final hybrid system. Therefore, the fermi level of the TMDCs must have been lower than the QDs before the hybrid device formed. This indicates that an electric field in the opposite direction as in the case of $ZnI_2$+MPA-based-QDs is formed. The authors conclude that this reverse electric field, in combination with worse trap state suppression[10] are the reasons for low performance of the EDT-based hybrid devices.

Figure S7c shows that the noise levels follow a similar behavior as the $ZnI_2$+MPA treated devices. With the measured responsivities and noise levels, detectivities (D*) at the exciton wavelength is $3.4 \times 10^{10}$ Jones for $WS_2$ based device whereas for the $MoS_2$ based devices D* is calculated to be $2 \times 10^{8}$ Jones.

Regarding the speed of the devices Figure S7f shows that $WS_2$ based device is slower than the $MoS_2$ based device with time constants of 2.44 seconds to 0.8 seconds, similar to the $ZnI_2$+MPA treated devices. This can be due to poor trap state suppression in the $WS_2$ based device as in the devices presented in the main text.

## Statistics on Hybrid Devices

We have fabricated and measured several devices following the same procedure as described in the main text to test the reproducibility and the performance range of our hybrid photodetectors. These results are summarized in Table S2.

| | Device Number | Max. Responsivity (A $W^{-1}$) | Noise current – $i_n$ (A $Hz^{-1/2}$) | Detectivity (Jones) | Average Time Constant (s) |
|---|---|---|---|---|---|
| $WS_2$ based Devices | 1 | 1442 | $1 \times 10^{-12}$ | $1.02 \times 10^{12}$ | 0.20 ± 0.11 |
| | 2 | 2530 | $1 \times 10^{-12}$ | $1.26 \times 10^{12}$ | 0.27 ± 0.05 |
| | 3 | 890 | $1.74 \times 10^{-12}$ | $3.25 \times 10^{11}$ | 0.45 ± 0.10 |
| | 4 | 1078 | $1 \times 10^{-12}$ | $7.62 \times 10^{11}$ | 0.36 ± 0.13 |
| $MoS_2$ based Devices | 1 | 202 | $5 \times 10^{-12}$ | $2.8 \times 10^{11}$ | 0.0296 ± 0.0264 |
| | 2 | 6.3 | $1.2 \times 10^{-12}$ | $3.7 \times 10^{9}$ | 0.0941 |
| | 3 | 1.01 | $5.1 \times 10^{-12}$ | $1.4 \times 10^{8}$ | 0.14 ± 0.017 |
| | 4 | 5.4 | $1.1 \times 10^{-12}$ | $3.47 \times 10^{9}$ | 0.062 ± 0.03 |

Table S2. A summary of the performance metrics of fabricated hybrid photodetectors.

All of the devices in Table S2 are fabricated on separate substrates and measured under the same conditions as the devices presented in the main text. In general, we can see that the responsivity values of $WS_2$ based devices are larger than the ones of $MoS_2$, which is the major difference between the two sets of devices towards a large difference in detectivities. Noise levels are almost the same in both cases and they are measured by taking the FFT of the dark current traces and confirmed by lock-in measurements. The speed of the detectors is measured using a 1310 nm laser, as in the case of the devices in the main text. Several measurements have been made with varying gate voltages for the speed and an average is established with an error. The results indicate that $WS_2$ based devices are generally slower than the $MoS_2$ based ones.

**State of the Art TMDC – CQD Hybrid Photodetectors**

| Active Material | Reference | Spectral Coverage (nm) | Responsivity ($AW^{-1}$) | Detectivity (Jones) | Decay Time (s) |
|---|---|---|---|---|---|
| $MoS_2$/PbS | [11] | 400-1500 | $6 \times 10^5$ | $5 \times 10^{11}$ | 0.3-0.4 |
| $MoS_2$/HgTe | [12] | Vis-2100 | $\sim 10^6$ | $\sim 10^{12}$ | $4 \times 10^{-3}$ |
| $WS_2$/PbS | [13] | 400-1500 | 14 | $3.9 \times 10^8$ | $2 \times 10^{-4}$ |
| $WSe_2$/PbS | [14] | 400-1400 | $2 \times 10^5$ | $7 \times 10^{13}$ | $7 \times 10^{-3}$ |
| $MoS_2$/graphene QDs | [15] | 400-700 | $1.6 \times 10^4$ | not reported | 1.23 |
| $MoS_2$/PbS | This work | 800-2100 | 202 | $2.8 \times 10^{11}$ | 0.03 |
| $WS_2$/PbS | This work | 800-2100 | 1442 | $1.02 \times 10^{12}$ | 0.2 |

Table S3. Previously reported hybrid devices consisting of TMDCs and quantum dots and some of their performance metrics.

## SUPPORTING INFORMATION REFERENCES

(1) Moreels, I.; Lambert, K.; Muynck, D. De; Vanhaecke, F.; Poelman, D.; Martins, J. C.; Allan, G.; Hens, Z. Size-Dependent Optical Properties of Colloidal PbS Quantum Dots. *ACS Nano* **2009**, *3* (10), 3023–3030.

(2) Bi, Y.; Bertran, A.; Gupta, S.; Ramiro, I.; Pradhan, S.; Christodoulou, S.; Majji, S.-N.; Akgul, M. Z.; Konstantatos, G. Solution Processed Infrared- and Thermo-Photovoltaics Based on 0.7 EV Bandgap PbS Colloidal Quantum Dots. *Nanoscale* **2019**, *11* (3), 838–843.

(3) Chiu, M. H.; Tseng, W. H.; Tang, H. L.; Chang, Y. H.; Chen, C. H.; Hsu, W. T.; Chang, W. H.; Wu, C. I.; Li, L. J. Band Alignment of 2D Transition Metal Dichalcogenide Heterojunctions. *Adv. Funct. Mater.* **2017**, *27* (19).

(4) Kozawa, D.; Carvalho, A.; Verzhbitskiy, I.; Giustiniano, F.; Miyauchi, Y.; Mouri, S.; Castro Neto, A. H.; Matsuda, K.; Eda, G. Evidence for Fast Interlayer Energy Transfer in MoSe2/WS2 Heterostructures. *Nano Lett.* **2016**, *16* (7), 4087–4093.

(5) Mak, K. F.; Lee, C.; Hone, J.; Shan, J.; Heinz, T. F. Atomically Thin MoS2: A New Direct-Gap Semiconductor. *Phys. Rev. Lett.* **2010**, *105* (13), 2–5.

(6) Zhao, W.; Ghorannevis, Z.; Chu, L.; Toh, M.; Kloc, C.; Tan, P.-H.; Eda, G. Evolution of Electronic Structure in Atomically Thin Sheets of WS2 and WSe2. *ACS Nano* **2013**, *7* (1), 791–797.

(7) Koh, W. K.; Koposov, A. Y.; Stewart, J. T.; Pal, B. N.; Robel, I.; Pietryga, J. M.; Klimov, V. I. Heavily Doped N-Type PbSe and PbS Nanocrystals Using Ground-State Charge Transfer from Cobaltocene. *Sci. Rep.* **2013**, *3*, 1–8.

(8) Kang, I.; Wise, F. W. Electronic Structure and Optical Properties of PbS and PbSe Quantum Dots. *J. Opt. Soc. Am. B* **1997**, *14* (7), 1632.

(9) Brown, P. R.; Kim, D.; Lunt, R. R.; Zhao, N.; Bawendi, M. G.; Grossman, J. C.; Bulović, V. Energy Level Modification in Lead Sulfide Quantum Dot Thin Films through Ligand Exchange. *ACS Nano* **2014**, *8* (6), 5863–5872.

(10) Pradhan, S.; Stavrinadis, A.; Gupta, S.; Bi, Y.; Di Stasio, F.; Konstantatos, G. Trap-State Suppression and Improved Charge Transport in PbS Quantum Dot Solar Cells with Synergistic Mixed-Ligand Treatments. *Small* **2017**, *13* (21), 1700598.

(11) Kufer, D.; Nikitskiy, I.; Lasanta, T.; Navickaite, G.; Koppens, F. H. L.; Konstantatos, G. Hybrid 2D-0D MoS2 -PbS Quantum Dot Photodetectors. *Adv. Mater.* **2015**, *27* (1), 176–180.

(12) Huo, N.; Gupta, S.; Konstantatos, G. MoS2 -HgTe Quantum Dot Hybrid Photodetectors beyond 2 Mm. *Adv. Mater.* **2017**, *29* (17), 1606576.

(13) Yu, Y.; Zhang, Y.; Song, X.; Zhang, H.; Cao, M.; Che, Y.; Dai, H.; Yang, J.; Zhang, H.; Yao, J. PbS-Decorated WS2 Phototransistors with Fast Response. *ACS Photonics* **2017**, *4* (4), 950–956.

(14) Hu, C.; Dong, D.; Yang, X.; Qiao, K.; Yang, D.; Deng, H.; Yuan, S.; Khan, J.; Lan, Y.; Song, H.; Tang, J. Synergistic Effect of Hybrid PbS Quantum Dots/2D-WSe2 Toward High Performance and Broadband Phototransistors. *Adv. Funct. Mater.* **2017**, *27* (2).

(15) Chen, C.; Qiao, H.; Lin, S.; Man Luk, C.; Liu, Y.; Xu, Z.; Song, J.; Xue, Y.; Li, D.; Yuan, J.; Yu, W.; Pan, C.; Ping Lau, S.; Bao, Q. Highly Responsive MoS2 Photodetectors Enhanced by Graphene Quantum Dots. *Sci. Rep.* **2015**, *5*, 1–9.